\documentclass[aps,superscriptaddress,showpacs,pra,amssymb, amsmath]{revtex4}
\usepackage{amsmath,latexsym,amssymb}
\newtheorem{lem}{Lemma}
\newtheorem{thm}{Theorem}

\def\keywords{\vspace{-.3em}
    \if@twocolumn
      \small\it Keywords\/\bf---$\!$%
    \else
      \begin{center}\small\bf Keywords\end{center}\quotation\small
    \fi}
\def\endkeywords{\vspace{0.6em}\par\if@twocolumn\else\endquotation\fi
    \normalsize\rm}
\def\appendix{\par
    \setcounter{section}{0}\setcounter{subsection}{0}
    \def\thesection{\Alph{section}} \section*{Appendix}
}
\def\Tr{\mathop{\rm Tr}\nolimits}

\def\SU{\mathop{\rm SU}\nolimits}

\def\real{\mathbb{R}}
\def\complex{\mathbb{C}}
\def\integer{\mathbb{Z}}

\def\Label#1{\label{#1}\ [\ #1\ ]\ }
\def\Bibitem#1{\bibitem{#1}\ [\ #1\ ]\ }
\def\Label{\label}
\def\Bibitem{\bibitem}
\begin{document}
\title{Quantum universal variable-length source coding}
\author{Masahito Hayashi}
\email{masahito@brain.riken.go.jp}
\affiliation{Laboratory for Mathematical Neuroscience, Brain Science Institute, RIKEN,
2-1 Hirosawa, Wako, Saitama, 351-0198, Japan}
\author{Keiji Matsumoto}
\email{keiji@qci.jst.go.jp}
\affiliation{Quantum Computation and Information Project, ERATO, JST,
5-28-3, Hongo, Bunkyo-ku, Tokyo, 113-0033, Japan}

\date{5 June 2002}
%03.67.-a    Quantum information
%03.67.Dd   Quantum cryptography 
%03.67.Hk   Quantum communication 
%03.67.Lx   Quantum computation 
\begin{abstract}
We construct an optimal quantum universal variable-length code
that achieves the admissible minimum rate,
i.e., our code is used for any probability 
distribution of quantum states.
Its probability of exceeding the admissible minimum rate
exponentially goes to $0$.
Our code is optimal in the sense of its exponent.
In addition, its average error 
asymptotically tends to $0$.
\end{abstract}
\maketitle

\section{Introduction}
As was proven by Schumacher \cite{Schumacher}, and Jozsa and Schumacher \cite{JS},
we can compress the unknown source state
into the length $n H(\overline{\rho}_p)$
with a sufficiently small error
when the source state on $n$ quantum systems 
obeys the $n$-i.i.d. distribution
of the known probability $p$,
where $\overline{\rho}_p:= \sum_{\rho} p(\rho) \rho$
and $H(\rho)$ is the von Neumann entropy $- \Tr \rho
\log \rho$.
Jozsa and Schumacher's protocol depends on the mixture state $\overline{\rho}_p$.
Concerning the quantum source coding,
there are two settings:
blind coding,
in which the input is the unknown quantum state,
and visible coding,
in which the input is classical information
which determines the quantum state that we want to send,
i.e., the encoder knows the input quantum state.
In this paper, we treat only blind coding.
In our setting, we allow mixed states as input states.

In blind coding,
Koashi and Imoto \cite{KI} proved that
even if we allow mixed states as input states
without trivial redundancies,
the minimum admissible length 
is $n H(\overline{\rho}_p)$.
Depending only on the coding length $n R$,
Jozsa et al. \cite{JH}
constructed a code
which is independent of 
the distribution which the input obeys.
In their protocol, 
if and only if the minimum admissible length of 
the distribution $p$ is less than
$n R$, 
we can decode with a sufficiently small error.
This kind code is called
a quantum universal fixed-length source code.

In the classical system,
depending on the input state,
the encoder can determine the coding length.
Such a code is called a variable-length code.
Using this type code, we can compress any information 
without error.
When we suitably choose a variable-length 
code for the probability distribution $p$ of the input,
the coding length is less than
$n H(p)$, except for a small enough probability.
In particular, Lynch \cite{Ly} and Davisson \cite{Da}
proposed a variable-length code with no error,
in which the coding length is less than
$n H(p)$ except for a small enough probability under 
the distribution $p$.
Such a code is called
a universal variable-length source code.
Today, their code can be regarded 
as the following two-stage code:
at the first step, we send the empirical distribution (i.e., the type)
which indicates a subset of data,
and in the second step, we send information
which indicates every sequence belonging to the subset \cite{CK}.

%In this paper,
%in the quantum case, we treat codes
This paper deals with 
quantum data compression 
in which the encoder determines the coding length, 
according to the input state.
In order to make this decision, he must measure the input 
quantum system.
After this measurement,
depending on the data, the encoder 
compresses the final state of this measurement
and sends its data and the compressed state.
This type code is called a {\it quantum variable-length source code}.
However, in general, 
the encoder knows only that
the input state is written as a separable state
$\rho_{x_1}\otimes \rho_{x_2}
\otimes \cdots \otimes \rho_{x_n}$.
Therefore, it is impossible to 
determine the coding length without 
destruction of the input state.

In particular, independently of the probability distribution $p$,
we construct the code satisfying 
the following conditions:
the average error 
concerning to Bures distance tends to $0$.
The probability that the coding length is 
greater than $n H(\overline{\rho}_p)$,
tends to $0$.
Such a code is called 
a {\it quantum universal variable-length source code}.
In our construction,
similarly to Keyl and Werner \cite{KW},
an essential role is played by the representation theory of 
the special unitary group and the symmetric group on the 
tensored space.
In our code, 
the encoder performs a quantum measurement 
closely related to irreducible decomposition
of the two groups, 
and its resulting data can be approximately regarded as
a quantum analogue of type.
Thus, our code can be regarded as
a quantum analogue of Lynch-Davisson code \cite{Ly},\cite{Da}.
Of course, if we can estimate the entropy 
$H(\overline{\rho}_p)$, we can compress
the coding rate to the admissible rate
$H(\overline{\rho}_p)$ with a probability close to 1.
However, when we perform a naive measurement for the
estimation of $H(\overline{\rho}_p)$,
the input state is destroyed.
Therefore, in our code,
it is the main problem to treat the trade-off
between the estimation of $H(\overline{\rho}_p)$
and the non-demolition of the input state.
%As discussed in section \ref{s6}, 

One might consider that the universal variable 
code can be easily realized as follows.
First, use the $n \epsilon$, where $\epsilon$ is small,
states for the estimation of $H(\overline{\rho}_p)$.
Second, apply Jozsa et. al. protocol \cite{JH} by 
setting $R=H(\overline{\rho}_p)+ \epsilon $,
and apply to $n(1-\epsilon)$ states.
If we consider individual error (\ref{stst}),
this code successfully compress the source.
However, in our paper, like Jozsa et. al. \cite{JH},
we consider the total Bures distance (\ref{st})
between the input state and the output state.
In this criterion, `naive estimate and compress'
strategy destroys the input state a lot.
The detail will be discussed in section 6.
(Note also that our criterion (\ref{st})
is different from Krattenthanler and Slater's 
criterion \cite{KS}
and Schumacher and Westmoreland's criterion \cite{SW}.)

In this paper, 
we discuss the universality for 
the probability family ${\cal P}$ 
consisting of predicted probabilities on
${\cal S}({\cal H})$.
For any probability family ${\cal P}$ on
${\cal S}({\cal H})$,
we define universality of a 
quantum variable-length source code
and evaluate the exponent of 
the probability that
the coding length is greater than 
the minimum admissible length,
which is called the overflow probability.
However, unfortunately,
in our approach, it is difficult to construct 
a quantum universal variable-length source code
whose error exponentially tends to $0$ 
in the blind setting.
In the visible coding case,
it is possible to construct 
such a code. 
This topic will be discussed in another paper.

We summarize quantum fixed-length source coding
in section \ref{s2.1}.
After this summary,
we state our mathematical setting 
and the main results in section \ref{s2.2}.
Our proofs and our construction of
code are given in sections \ref{s3} and \ref{s4}.
Moreover, 
as is demonstrated in section \ref{s6}, in the 2-dimensional case, 
a naive code destroys the state
and is not used as a quantum universal variable-length
source code.

%\section{Main result}\Label{s2}

\section{Summary of quantum fixed-length source coding}\Label{s2.1}
Let ${\cal H}$ be a finite-dimensional Hilbert space 
that represents the physical system of interest
and let ${\cal S}({\cal H})$ be the set of
density operators on ${\cal H}$.
Consider a source of quantum state which produces
the state $\vec{\rho}_n
:=\rho_1 \otimes \rho_2 \otimes \cdots \otimes\rho_n$
with probability the i.i.d. distribution $p^n$
of the probability $p$ on ${\cal S}({\cal H})$.
In {\it fixed-length source coding},
a sequence of states 
$\vec{\rho}_n$
is compressed 
to the state in a smaller Hilbert space 
${\cal H}_n \subset {\cal H}^{\otimes n}$,
whose dimension is $e^{nR}$.
Here, the encoder and the decoder is
a trace-preserving completely positive 
(TP-CP) map $E^n$ and $D^n$, respectively. 
The average of the total error is given by
\begin{align}
\epsilon_{n,p}(E^n, D^n):=
\sum_{\vec{\rho}_n \in {\cal S}({\cal H}^{\otimes n})}
p^n(\vec{\rho}_n)
b^2\left(\vec{\rho}_n,  
D^n \circ E^n( \vec{\rho}_n)
\right) , \Label{st}
\end{align}
where
Bures distance is defined as
\begin{align*}
b(\rho,\sigma):= \sqrt{
1- \Tr \left|
\sqrt{\rho}\sqrt{\sigma}\right|} .
\end{align*}
Note that the support of $p$ does not
necessarily consist of pure states.
In this setting, we focus the infimum of the rate 
with which the average error goes to zero.
The infimum is called the minimum admissible rate $R_p$ 
of $p$, and is defined by
\begin{align*}
R_p:=
\inf\left\{\left.
\limsup \frac{1}{n} \log \dim {\cal H}_n \right|
\exists \{ ({\cal H}_n, E^n,D^n) \}, \quad
\epsilon_{n,p}(E^n, D^n) \to 0 
\right\}.
\end{align*}
The number $nR_p$ is called minimum admissible length.
When the source has no trivial redundancy in the sense following,
it is calculated as
\begin{align*}
R_p= H( \overline{\rho}_p):= 
-\Tr \overline{\rho}_p\log \overline{\rho}_p,
\end{align*}
where $\overline{\rho}_p:=
\sum_{\rho \in {\cal S}({\cal H})}
p(\rho) \rho$.
The direct part was proven by
Schumacher \cite{Schumacher},and Jozsa and Schumacher \cite{JS},
and the converse part 
was proven by Barnum et al. \cite{Barnum}
in the pure state case.
In the mixed case,
Koashi and Imoto \cite{KI} 
discussed this problem
as follows.
Indeed, if the source has trivial redundancies,
we can compress up to more than the rate $H(\overline{\rho}_p)$.
We consider
the source to have trivial redundancy
if the support ${\cal S}(p)$ of $p$ satisfies the 
following.
The Hilbert space ${\cal H}$ is decomposed as
(\ref{3/26}) satisfying the conditions {\bf (i)} and 
{\bf (ii)}:
\begin{align}
{\cal H}= 
\bigoplus_l {\cal H}_{J,l} \otimes {\cal H}_{K,l}  \Label{3/26}
\end{align}
\begin{description}
\item{\bf (i)}
Any element $\rho \in {\cal S}(p)$
is commutative with $P_l $,
where
$P_l$ denotes the projection to the 
subspace ${\cal H}_{J,l} \otimes {\cal H}_{K,l}$.
\item{\bf (ii)}
The state $\frac{\Tr _{{\cal H}_{J,l}} P_l \rho P_l}{\Tr P_l \rho}$
is independent of $\rho\in {\cal S}(p)$.
\end{description}
Precisely, we should state that 
the conditions {\bf (i)} and 
{\bf (ii)} hold almost everywhere for $p$.
In this case, without loss of information,
we can transform 
$\rho$ to $\sum_l \Tr _{{\cal H}_{K,l}} P_l \rho P_l$.
When the encoder sends the state
$\sum_l \Tr _{{\cal H}_{K,l}} P_l \rho P_l$
instead of $\rho$,
the decoder can recover the state $\rho$ from the state
$\sum_l \Tr _{{\cal H}_{K,l}} P_l \rho P_l$.
This fact implies that
we can compress up to the rate
$H\left( \sum_{\rho} p(\rho) \sum_l 
\Tr _{{\cal H}_{K,l}} P_l \rho P_l\right)$,
i.e. 
$
R_p \le  H\left( \sum_{\rho} p(\rho) \sum_l 
\Tr _{{\cal H}_{K,l}} P_l \rho P_l\right)
$.
Koashi and Imoto also proved the opposite inequality,
i.e. proved 
the equation
\begin{align}
R_p= H\left( \sum_{\rho} p(\rho) \sum_l 
\Tr _{{\cal H}_{K,l}} P_l \rho P_l\right),\Label{D4-1}
\end{align}
where the RHS of (\ref{D4-1}) is given by the 
finest decomposition satisfying 
{\bf (i)} and {\bf (ii)}.
Following their proof, we can understand that
if $\limsup  \frac{1}{n} \log \dim {\cal H}_n 
\,< R_p$,
\begin{align}
\liminf \epsilon_{n,p}(E^n, D^n) \,> 0, \Label{D4}
\end{align}
which is called the weak converse.
When the support of $p$ consists of pure states,
if  $\limsup  \frac{1}{n} \log \dim {\cal H}_n 
\,< R_p= H(\overline{\rho}_p)$,
we obtain 
\begin{align}
\lim \epsilon_{n,p}(E^n, D^n) =1, \Label{D4-3}
\end{align}
which is called the strong converse,
and was proven by Winter \cite{Winter}
in the first time.
A more simple proof was given by Hayashi \cite{H2002}.
However, the strong converse in the mixed states case
is an open problem.
Moreover, in the pure states case,
the optimal exponent of average error 
was treated by Hayashi \cite{H2002}.
\section{Quantum universal variable-length source coding}
\Label{s2.2}
In the variable-length case,
we need describe a quantum measurement
with state evolution,
by using {\it an instrument} 
consisting of a decomposition ${\bf E}'= \{ {\bf E}'_\omega
\}_{\omega \in \Omega}$,
by CP maps 
from ${\cal S}({\cal H})$ to ${\cal S}({\cal H})$
under the condition 
$\sum_{\omega \in \Omega} \Tr {\bf E}'_\omega
(\rho) = 1 ,\quad \forall \rho \in {\cal S}({\cal H})$.
When we perform the instrument ${\bf E}'= 
\{ {\bf E}'_\omega\}_{\omega\in \Omega}$
for an initial state $\rho$,
we get the data $\omega$ and the final state 
$\frac{{\bf E}'_\omega(\rho)}
{\Tr {\bf E}'_\omega(\rho)}$ with the
probability $\Tr {\bf E}'_\omega(\rho)$.
A quantum variable-length encoder ${\bf E}$ is 
given by 
a measurement process ${\bf E}'$ and
encoding process $E_{\omega}''$ depending the data $\omega$,
which is 
a TP-CP map from ${\cal S}({\cal H})$ to 
${\cal S}({\cal H}_{\omega})$,
where the Hilbert space ${\cal H}_{\omega}$ depends
on the data $\omega$, as
\begin{align*}
{\bf E}_{\omega}=  E_{\omega}''\circ {\bf E}'_{\omega}.
\end{align*}
Therefore,
any quantum variable-length encoder ${\bf E}$ consists
of 
a decomposition ${\bf E}= \{ {\bf E}_\omega
\}_{\omega \in \Omega}$,
by CP maps 
from ${\cal S}({\cal H})$ to ${\cal S}({\cal H}_\omega)$
under the condition 
$\sum_{\omega \in \Omega} \Tr {\bf E}_\omega
(\rho) = 1 ,\quad \forall \rho \in {\cal S}({\cal H})$.
For a detail about instruments, see Ozawa 
\cite{Oz1,Oz2,Oz3}.

The decoder is given by 
a set of TP-CP maps ${\bf D} = \{ {\bf D}_{\omega}\}_{
\omega \in \Omega}$,
which presents the decoding process depending
the data $\omega$.
A pair of an encoder ${\bf E} = \{ 
{\bf E}_{\omega} \}_{\omega \in \Omega}$ and 
a decoder ${\bf D} = \{ {\bf D}_{\omega}\}_{
\omega \in \Omega}$
is called 
a {\it quantum variable-length source code}
on ${\cal H}$.
The coding length is described by
$\log | \Omega | + \log \dim {\cal H}_{\omega}$,
which is a random variable
obeying the probability
${\rm P}_{\rho}^{{\bf E}}(\omega)
:=
\Tr {\bf E}_{\omega}(\rho)$
when the input state is $\rho$.
Of course, any quantum variable-length
source code can be regarded as
a quantum fixed-length source code whose
length is the maximum of 
$\log | \Omega | + \log \dim {\cal H}_{\omega}$.

When the state $\vec{\rho}_n$ on 
${\cal H}^{\otimes n}$
obeys the i.i.d. 
distribution $p^n$ of the probability
$p$ on 
${\cal S}({\cal H})$,
the error of decoding for 
a variable-length code $({\bf E}^n,{\bf D}^n)$
on ${\cal H}^{\otimes n}$
is
evaluated by Bures distance as 
\begin{align*}
\sum_{\omega_n \in \Omega_n}
\Tr {\bf E}^n_{\omega_n}(\vec{\rho}_n)
b^2\left(\vec{\rho}_n,  
D_{\omega_n} \left( 
\frac{{\bf E}^n_{\omega_n}(\vec{\rho}_n)}
{\Tr {\bf E}^n_{\omega_n}(\vec{\rho}_n)}
\right) \right), 
\end{align*}
and 
the average error is given by
\begin{align}
\epsilon_{n,p}({\bf E}^n,{\bf D}^n):=
\sum_{\vec{\rho}_n \in {\cal S}({\cal H}^{\otimes n})}
p^n(\vec{\rho}_n)
\sum_{\omega_n \in \Omega_n}
\Tr {\bf E}^n_{\omega_n}(\vec{\rho}_n)
b^2\left(\vec{\rho}_n,  
D_{\omega_n} \left( 
\frac{{\bf E}^n_{\omega_n}(\vec{\rho}_n)}
{\Tr {\bf E}^n_{\omega_n}(\vec{\rho}_n)}
\right) \right).
\Label{e90}
\end{align}
In this case, 
the data $\omega_n$ obeys the probability:
\begin{align}
{\rm P}_{p^n}^{{\bf E}^n}(\omega_n)
:=
\sum_{\vec{\rho}_n \in {\cal S}({\cal H}^{\otimes n})}
p^n(\vec{\rho}_n)
\Tr {\bf E}^n_{\omega_n}(\vec{\rho}_n)
=
\Tr {\bf E}^n_{\omega_n}(\overline{\rho}_p^{\otimes n})
. \Label{4-13}
\end{align}
A sequence $\{({\bf E}^n,{\bf D}^n)\}$
of 
quantum variable-length source code
is called {\it universal}
for a probability family ${\cal P}$ on ${\cal S}({\cal H})$
if 
\begin{align*}
\epsilon_{n,p}({\bf E}^n,{\bf D}^n) \to 0 
\end{align*}
for any probability $p \in {\cal P}$.
%In particular,
%in the case of all probabilities on ${\cal S}({\cal H})$,
%we call it universal.

As guaranteed by Theorem \ref{t2},
we can reduce the coding rate to the admissible 
rate $H(\overline{\rho}_p)$ with a sufficiently 
small error and a probability infinitely close to $1$,
asymptotically, i.e.
there exists a quantum universal 
variable-length source code
$\{({\bf E}^{n}, {\bf D}^{n})\}$ satisfying that 
\begin{align}
\lim
{\rm P}_{p^n}^{{\bf E}^n}
\left\{ \frac{1}{n}
\left( \log | \Omega_n | + \log \dim {\cal H}_{\omega_n}
\right) \ge H(\overline{\rho}_p) +\epsilon
 \right\}
= 0,
\quad
\forall \epsilon \,> 0, \forall p \in {\cal P}.
\Label{4-13-3}
\end{align}
Conversely, if 
a quantum variable-length source code
$\{({\bf E}^{n}, {\bf D}^{n})\}$
is universal for a family ${\cal P}$ and
\begin{align}
\lim
{\rm P}_{p^n}^{{\bf E}^n}
\left\{ \frac{1}{n}
\left( \log | \Omega_n | + \log \dim {\cal H}_{\omega_n}
\right) \ge R
 \right\}
= 0, \Label{4-13-2}
\end{align}
then $R \ge R_p$
because the inequality (\ref{4-13-2})
implies the existence of a fixed-length code
with the rate $R$ and an asymptotically small error.
When two probabilities $p, q \in {\cal P}$ satisfy
that $\overline{\rho}_p= \overline{\rho}_q$,
equation (\ref{4-13}) guarantees that
${\rm P}_{p^n}^{{\bf E}^n}={\rm P}_{q^n}^{{\bf E}^n}$.
Thus,
any quantum universal 
variable-length source code
$\{({\bf E}^{n}, {\bf D}^{n})\}$
satisfies the inequality
\begin{align*}
\inf\left\{ R \left|
\lim
{\rm P}_{p^n}^{{\bf E}^n}
\left\{ \frac{1}{n}
\left( \log | \Omega_n | + \log \dim {\cal H}_{\omega_n}
\right) \ge R
 \right\}
= 0 \right.\right\} 
\ge \sup_{q\in {\cal P}:
\overline{\rho}_p= \overline{\rho}_q}
R_q.
\end{align*}
Therefore,
the inequalities
\begin{align}
H(\overline{\rho}_p)
&\ge
\sup_{\{({\bf E}^{n}, {\bf D}^{n})\}:
\hbox{\footnotesize univ. for } {\cal P}}
\inf\left\{ R \left|
\lim
{\rm P}_{p^n}^{{\bf E}^n}
\left\{ \frac{1}{n}
\left( \log | \Omega_n | + \log \dim {\cal H}_{\omega_n}
\right) \ge R
 \right\}
= 0 \right.\right\} \nonumber \\
&\ge \sup_{q\in {\cal P}:
\overline{\rho}_p= \overline{\rho}_q}
R_q \Label{op9}
\end{align}
hold.
When the support of $p$ consists of
pure states,
since the admissible rate $R_p$ equals 
$H(\overline{\rho}_p)$,
the RHS of (\ref{op9}) equals $H(\overline{\rho}_p)$
i.e. our code is optimal.
However, in the mixed states case, 
the admissible rate $R_p$ of a probability
$p$ is rarely less than $H(\overline{\rho}_p)$.
(See equation (\ref{D4-1}).)
In this rare case, our code cannot
up to the admissible rate $R_p$.
When for any $\rho \in {\cal S}({\cal H})$
there exists a probability $q \in {\cal P}$ such that
$\overline{\rho}_q= \rho$ and
$R_q=H(\overline{\rho}_q)$,
the RHS of (\ref{op9}) equals $H(\overline{\rho}_p)$,
although
the admissible rate $R_p$ 
is less than $H(\overline{\rho}_p)$.
In this case, our code is optimal
for any probability $p \in {\cal P}$.

Next, we discuss the exponent of 
the overflow probability:
${\rm P}_{p^n}^{{\bf E}^n}
\left\{ \frac{1}{n}
\left( \log | \Omega_n | + \log \dim {\cal H}_{\omega_n}
\right) \ge R \right\}$.
\begin{thm}\Label{t2}
For any family ${\cal P}$,
there exists a quantum variable-length source code
$\{({\bf E}^{n}, {\bf D}^{n})\}$ 
on ${\cal H}^{\otimes n}$
which satisfies the condition that
$\epsilon_{n,p}({\bf E}^{n}, {\bf D}^{n})$
tends to $0$ uniformly for $p \in {\cal P}$
and that
\begin{align}
\lim
\frac{-1}{n}\log 
{\rm P}_{p^n}^{{\bf E}^n}
\left\{ \frac{1}{n}
\left( \log | \Omega_n | + \log \dim {\cal H}_{\omega_n}
\right) \ge R \right\}
=
\inf_{q \in {\cal P} : H(\overline{\rho}_q)\ge R } 
\min_{V:\hbox{\footnotesize unitary}}
D(\overline{\rho}_q \| V \overline{\rho}_p V^*), 
\Label{op3}
\end{align}
where $D(\rho\|\sigma)$ 
is quantum relative entropy $\Tr \rho (\log \rho - \log \sigma)$.
\end{thm}
Of course, when
the set ${\cal S}:=
\{ \overline{\rho}_p |p \in {\cal P}\}$
is unitary invariant,
the RHS equals
$\inf_{q \in {\cal P} : H(\overline{\rho}_q)\ge R } 
D(\overline{\rho}_q \| \overline{\rho}_p )$.
We construct a quantum variable-length source code
satisfying (\ref{op3}) in section \ref{s4}.
Indeed, as is guaranteed by the following theorem,
our code is optimal in the sense of
the exponent of the decreasing rate of 
the overflow probability
when $\inf_{q \in {\cal P} : 
H(\overline{\rho}_q)\ge R } 
\min_{V:unitary}
D(\overline{\rho}_q \| V \overline{\rho}_p V^*)=
 \inf_{q \in {\cal P}: R_q \,> R}
D(\overline{\rho}_q \| \overline{\rho}_p )$.
\begin{thm}\Label{t1}
If a sequence $\{({\bf E}^n,{\bf D}^n)\}$
of quantum variable-length source codes
on ${\cal H}^{\otimes n}$
is universal for a family ${\cal P}$,
then
\begin{align}
\limsup
\frac{-1}{n}\log 
{\rm P}_{p^n}^{{\bf E}^n}
\left\{ \frac{1}{n}
\left( \log | \Omega_n | + \log \dim {\cal H}_{\omega_n}
\right) \ge R \right\}
\le 
\inf_{q \in {\cal P}: R_q \,> R}
D(\overline{\rho}_q \| \overline{\rho}_p ). \Label{op}
\end{align}
\end{thm}
Of course, when the family consists 
of all probabilities on ${\cal S}({\cal H})$,
the RHS of (\ref{op3}) and the RHS of (\ref{op})
coincide, i.e. our code is optimal in the sense
of the exponent of the overflow probability.

\section{Construction of a 
quantum variable-length source code}\Label{s4}
%Proof of Theorem \ref{t2}
First, we construct
a universal quantum variable-length source code
that achieves the optimal rate
(\ref{op3})
for the family of all probabilities on ${\cal S}({\cal H})$.
This family is covariant
for the actions of the $d$-dimensional special unitary group $\SU(d)$,
and any $n$-i.i.d. distribution $p^n$
is invariant for the action of the $n$-th symmetric group $S_n$
on the tensored space ${\cal H}^{\otimes n}$.
Thus, our code should satisfy the invariance
for these actions on ${\cal H}^{\otimes n}$.

Now, we focus on the irreducible decomposition 
of the tensored space ${\cal H}^{\otimes n}$
concerning the representations of 
$S_n$ and $\SU(d)$,
and define the Young index ${\bf n}$ as,
\begin{align*}
{\bf n}: = (n_1, \ldots, n_d) , \quad \sum_{i=1}^d
n_i= n, n_{i} \ge n_{i+1},
\end{align*}
and denote the set of 
Young indices ${\bf n}$ by $Y_n$.
Young index ${\bf n}$ uniquely corresponds to 
the irreducible unitary representation 
of $S_n$ and
the one of $\SU(d)$.
Now, we denote the representation space
of the irreducible unitary representation 
of $S_n$ ($\SU(d)$) corresponding to ${\bf n}$
by ${\cal V}_{{\bf n}}$ (${\cal U}_{{\bf n}}$), respectively.
In particular, regarding a unitary representation 
of $\SU(d)$,
Young index ${\bf n}$ gives the highest weight of 
the corresponding representation.
Then, the tensored space ${\cal H}^{\otimes n}$ is 
decomposed as follows; i.e.
${\cal H}^{\otimes n}$ is equivalent with the following
direct sum space under the representation of
$S_n$ and $\SU(d)$.
\begin{align*}
{\cal H}^{\otimes n}=
\bigoplus_{{\bf n}}{\cal W}_{{\bf n}} , \quad
{\cal W}_{{\bf n}}:= {\cal U}_{{\bf n}} \otimes {\cal V}_{{\bf n}} .
\end{align*}
For details, see Weyl \cite{Weyl}, 
Goodman and Wallach \cite{GW}, and Iwahori \cite{Iwa}.
The efficiency of this representation method
was discussed from several viewpoints.
Regarding fixed-length source coding,
it was discussed by
Jozsa et. al. \cite{JH}.
Regarding quantum relative entropy,
it was by Hayashi\cite{H1997}.
Regarding quantum hypothesis testing,
it wsa by Hayashi\cite{H2001}.
Regarding estimation of spectrum,
it was by Keyl and Wener\cite{KW}.

In the following, for an intuitive explanation of
our construction, 
we naively construct a good variable-length code
in the case ${\cal H}=\complex^2$.
For this construction,
we fixed a strictly increasing sequence 
$\vec{a}:=\{ a_i\}_{i=1}^{l+1}$
of real numbers
such that
$\frac{1}{2}
=a_1 \,< a_2 \,< \ldots \,< a_l \,< a_{l+1}=1$.
We define the encoder ${\bf E}^{\vec{a},n}$
whose data set $\{ 1, \ldots, l\}$ by
\begin{align*}
{\cal H}^{\vec{a},n}_i
&:= \oplus_{\frac{\bf n}{n} :
\frac{n_1}{n} \in [  a_i,a_{i+1}) } {\cal W}_{\bf n}\quad i = 1 \ldots l-1 \\
{\cal H}^{\vec{a},n}_l
&:= \oplus_{\frac{\bf n}{n} :
\frac{n_1}{n} \in [  a_l,a_{l+1}] } {\cal W}_{\bf n}\\
{\bf E}_i^{\vec{a},n}(\rho_n)
&:=
P^{\vec{a},n}_i  \rho_n P^{\vec{a},n}_i ,\quad
\rho_n \in {\cal S}({\cal H}).
\end{align*}
and define the decoder ${\bf D}_i^{\vec{a},n}$
as the embedding from ${\cal H}^{\vec{a},n}_i$
to ${\cal H}^{\otimes n}$,
where we denote 
the projection to ${\cal H}^{\vec{a},n}_i$ by
$P^{\vec{a},n}_i$.
When the larger eigenvalue of the mixture $\overline{\rho}_p$
belongs to the interval $[  a_i,a_{i+1})$,
as is guaranteed by Lemma \ref{lee} in Appendix \ref{asa},
if the larger eigenvalue of the mixture $\overline{\rho}_p$ 
does not
lie on the boundary on the interval $[  a_i,a_{i+1})$,
the probability $\Tr \overline{\rho}_p^{\otimes n}
P^{\vec{a},n}_i $ tends to $1$.
Thus, we can prove $\epsilon_{n,p}(
{\bf E}^{\vec{a},n},{\bf D}^{\vec{a},n})  \to 0$.
Its speed depends on the divergence between 
the probability and the boundary.
Of course, if we choose $a_{i+1}-a_i$ to be sufficiently
small,
the coding length is close to the entropy 
$H(\overline{\rho}_p)$
with almost probability $1$.
However, when the larger eigenvalue lies on 
the boundary, the state is demolished,
as is caused by the same reason of Lemma \ref{L1}.
In this case, similarly to Lemma \ref{L1},
we can prove
\begin{align*}
\lim \epsilon_{n,p}(
{\bf E}^{\vec{a},n},{\bf D}^{\vec{a},n})  \,> 0.
\end{align*}

Now, we assume that the interval $a_{i+1}- a_i ~(
i=2, \ldots , l-1)$ is $\delta$
and that $a_2-a_1, a_{l+1}-a_l \,< \delta$.
Then, our code is uniquely defined by the choice
of $a_2 \in (\frac{1}{2}, \frac{1}{2}+\delta)$.
For the non-demolition of initial states,
we construct a variable-length code,
by choosing $a_2 \in 
\{ \frac{k}{n}| \frac{k}{n} \in 
(\frac{1}{2}, \frac{1}{2}+\delta), k \in\mathbb{Z} \}$ 
at random.
In this protocol,
we can expect that the average error tends to $0$
for any probability $p$ on ${\cal S}(\complex^2)$.
Note that the set $\{ \frac{k}{n}| \frac{k}{n} \in 
(\frac{1}{2}, \frac{1}{2}+\delta), k \in\mathbb{Z} \}$
corresponds to the data set $\Omega_n$.
In order to achieve the rate $H(\overline{\rho}_p)$,
we need to choose the set $\Omega_n$ so that
$\frac{1}{n} \log | \Omega_n| \to 0$.
It is essential for our code
to restrict $a_2$ to this lattice $\{\frac{k}{n}| k \in \mathbb{Z}\}$.

Moreover, for a fixed number $n$, when $\delta$ is large,
the demolition of initial state seems small and
the coding length seems long.
Therefore, roughly speaking,
in this code for a finite number $n$, 
by choosing $\delta$, we can treat the trade off
between the coding length and the non-demolition
of the input state.

Next, we generalize the above code
to the $d$-dimensional case,
and evaluate its average error.
In order to satisfy the universality and
the condition (\ref{op3}),
we need choose $\delta$ depending on $n$, 
more carefully.
For $\delta \,> 0$ we define
a subset $Y_{\delta,n}$ of 
$\integer ^d$ as
\begin{align*}
Y_{\delta,n}:=
\left\{ {\bf k} \in \integer ^d \left|
\sum_{i=1}^d k_i=n, \exists {\bf n} 
\in Y_n \cap U_{{\bf k}, n \delta}
\right.\right\}, 
\end{align*}
and define an operator 
$M_{\bf k}^{\delta, n}$ for any element ${\bf k}
\in Y_{\delta,n}$ as 
\begin{align*}
M_{\bf k}^{\delta, n}&:=
\frac{1}{C_{1,d}(n\delta)}
P_{\bf k}^{\delta, n} \\
P_{\bf k}^{\delta, n}&:=
\sum_{{\bf n} \in Y_n\cap U_{{\bf k}, n \delta}} 
P_{{\bf n}} \\
U_{{\bf p}, \delta}&:=
\left\{\left.
{\bf q} \in \real^d \right| 
\| {\bf p}- {\bf q}\| \le \delta
\right\} \\
C_{1,d}(x) &:= 
\# \left\{ {\bf k} \in \integer^d \left| \|{\bf k}\| \le x ,
\sum_{i=1}^d k_i =0 \right.\right\}, 
\end{align*}
where
$P_{{\bf n}}$ denotes the projection to
${\cal W}_{{\bf n}}$.

The number 
$\#
\left\{
{\bf k} \in \integer^d \cap  U_{{\bf n}, n \delta}
\left|
\sum_{i=1}^d k_i = n
\right.
\right\}$
is independent for ${\bf n} \in Y_n$
and equals $C_{1,d}(n\delta)$.
Thus, we have the relations
\begin{align*}
P_{\bf n}
\sum_{{\bf k} \in Y_{\delta,n}} M_{\bf k}^{\delta, n}
P_{\bf n}
=
\frac{
\#
\left\{{\bf k} \in Y_{\delta,n}|
{\bf n } \in Y_n \cap U_{{\bf k},n \delta}\right\}
}{C_{1,d}(n\delta)}
P_{\bf n}
=P_{\bf n},
\end{align*}
which implies 
the condition
\begin{align*}
\sum_{{\bf k} \in Y_{\delta,n}} M_{\bf k}^{\delta, n}
= I .
\end{align*}
The encoder ${\bf E}^{\delta,n}$ whose data set
is $Y_{\delta,n}$
is defined by
\begin{align*}
{\cal H}_{\bf k}^{\delta ,n}& := 
\bigoplus _{{\bf n} \in Y_n: \| {\bf n} - {\bf k}\|
\le n \delta} {\cal W}_{{\bf n}} \\
{\bf E}^{\delta,n}_{{\bf k}}(\rho_n) &:=
\sqrt{M_{\bf k}^{\delta, n}} \rho_n
\sqrt{M_{\bf k}^{\delta, n}} ,
\quad \forall \rho_n \in 
{\cal S}({\cal H}^{\otimes n}),
\end{align*}
and 
the decoder ${\bf D}^{\delta,n}_{{\bf k}}$ is defined 
as the embedding from $
{\cal H}_{\bf k}^{\delta, n}$
to ${\cal H}^{\otimes n}$.

As is proven in Appendixes \ref{asb} and \ref{asc},
the quantum variable-length source code
$({\bf E}^{\delta,n}, {\bf D}^{\delta,n})$ 
on ${\cal H}^{\otimes n}$
satisfies
\begin{align}
&\epsilon_{n,p}({\bf E}^{\delta,n},{\bf D}^{\delta,n})
\le 
\inf_{\delta_1:0\,< \delta_1 \,< \delta}
1- 
\frac{C_{2,d}(n \delta_1)}{C_{1,d}(n \delta)}
\Bigl(
1- 
(n+d)^{4d }
\exp \left( -n 
C_{3,d} 
(\delta- \delta_1)^2
\right)
\Bigr)^{\frac{3}{2}} , 
\Label{e1} \\
&\frac{-1}{n}
\log {\rm P}_{p^n}^{{\bf E}^{\delta,n}}
\left\{ \frac{1}{n}
\left(\log | Y_{\delta,n} | 
+ \log \dim {\cal H}_{\bf k}^{\delta, n}
\right)
\ge R \right\} \nonumber \\
& \ge \frac{-5d}{n} \log (n+d)
+ \inf_{{\bf q}\in \real_+^{d,1}
: H({\bf q}) \ge R - \frac{4d}{n}\log (n+d)}
\inf_{{\bf q}'\in \real_+^{d,1}
: \| {\bf q} - {\bf q}' \|\le 2 \delta}
D( {\bf q}'\| {\bf p } ),
\Label{e2}
\end{align}
where
\begin{align}
C_{2,d}(x) &:= 
\min_{{\bf p} \in \real^d : \sum_i p_i =0}
\# \left\{ {\bf k} \in \integer^d \left| \|{\bf k}- 
{\bf p} \| \le x ,
\sum_{i=1}^d k_i =0 \right.\right\}, \nonumber\\
C_{3,d} & := \min_{{\bf q},{\bf p}\in \real_+^{d,1}}
\frac{D({\bf q}\|{\bf p})}{\| {\bf p} - {\bf q} \|^2} , 
\Label{a3}\\
\real_+& := \{ x \in \real | x \ge 0\}, \quad
\real_+^{d,1}:= \left\{ {\bf p} \in \real_+^d
\left| \sum_i p_i =1\right.\right\}
, \nonumber
\end{align}
and ${\bf p}\in \real_+^{d,1}$ denotes 
the probability $(p_1, p_2, \ldots, p_d)$,
where
$p_i $
is eigenvalue of $\overline{\rho}_p$ and
$p_1 \ge p_2 \ge \ldots \ge p_d$.
In this paper, we use an italic
alphabet $p$ for denoting a probability
on ${\cal S}({\cal H})$ while
we use a bold alphabet ${\bf p}$ for
denoting a probability
$(p_1, \cdots, p_d)$
on $\{ 1, \cdots, d\}$.
Note that the RHS of (\ref{e1}) is independent of $p$.
Our main point is 
simultaneously reducing
$\epsilon_{n,p}({\bf E}^{\delta,n},{\bf D}^{\delta,n})$
and  ${\rm P}_{p^n}^{{\bf E}^{\delta,n}}
\left\{ \frac{1}{n}
\left(\log | Y_{\delta,n} | 
+ \log \dim {\cal H}_{\bf k}^{\delta, n}
\right)
\ge R \right\}$.
The RHS of (\ref{e2}) decreases as 
$\delta $ increases while
the relation between the RHS of (\ref{e1}) 
and $\delta$ is not necessarily simple.
However, letting $\delta:= n^{-1/4}$ and
$\delta_1:= n^{-1/4} - n^{-1/3}$,
we can check that
the RHS of (\ref{e1}) tends to $0$, and that
the RHS of (\ref{e2}) tends to the RHS of (\ref{op3}).
Thus, we obtain Theorem \ref{t2} when ${\cal P}$ consists
of all probabilities on ${\cal S}({\cal H})$.

If we adopt another criterion:
\begin{align*}
\epsilon''_{n,p}({\bf E}^{n},{\bf D}^{n})
:=
\sum_{\vec{\rho}_n \in {\cal S}({\cal H}^{\otimes n})}
p^n(\vec{\rho}_n)
\sum_{\omega_n \in \Omega_n}
\Tr {\bf E}^n_{\omega_n}(\vec{\rho}_n)
\left( 1- 
\left( \Tr \left| \vec{\rho}_n 
D_{\omega_n} \left( 
\frac{{\bf E}^n_{\omega_n}(\vec{\rho}_n)}
{\Tr {\bf E}^n_{\omega_n}(\vec{\rho}_n)}
\right) \right| \right)^2
\right),
\end{align*}
we have the following inequality 
instead of (\ref{e1}):
\begin{align}
\epsilon_{n,p}''({\bf E}^{\delta,n},{\bf D}^{\delta,n})
\le 
\inf_{\delta_1:0\,< \delta_1 \,< \delta}
1- 
\frac{C_{2,d}(n \delta_1)}{C_{1,d}(n \delta)}
\Bigl(
1- 
(n+d)^{4d }
\exp \left( -n 
C_{3,d} 
(\delta- \delta_1)^2
\right)
\Bigr)^{2}  
\Label{e1-3} ,
\end{align}
which is proven in Appendix \ref{asc}.

Next, deforming the code
$({\bf E}^{\delta,n}, {\bf D}^{\delta,n})$,
we construct a universal quantum variable-length source code
that achieves the optimal rate
in the general case with no trivial redundancy.
Define the set $Y_{\delta,\delta_1,n}({\cal S})$
as
\begin{align*}
Y_{\delta,\delta_1,n}({\cal S}) :=
\left\{ {\bf k} \in Y_{\delta,n}\left|
\exists \rho \in {\cal S}, \quad
\left\| {\bf p}(\rho)- \frac{\bf k}{n}\right\| \le \delta_1 
\right.\right\} ,
\end{align*}
where ${\bf p}(\rho)$ consists eigenvalues 
of $\rho$ such that $p_1(\rho) \ge \ldots \ge
p_d(\rho)$.
In particular, ${\bf p}= {\bf p}(\overline{\rho}_p)$.
Note that ${\cal S}$ is defined 
after Theorem \ref{t2}, and
is different from ${\cal S}(p)$.
When the data ${\bf k}$ belongs to $Y_{\delta,\delta_1,n}({\cal S})$,
we send the state
$\frac{{\bf E}^{\delta,n}_{\bf k}
(\vec{\rho}_n)}
{\Tr {\bf E}^{\delta,n}_{\bf k}
(\vec{\rho}_n)}$.
Otherwise, we send only the classical information $0$,
except for $Y_{\delta,\delta_1,n}({\cal S})$.
Then, the data set of the encoder 
is $Y_{\delta,\delta_1,n,+}({\cal S}):=
Y_{\delta,\delta_1,n}({\cal S}) \cup \{ 0 \}$.
The decoder is defined as
\begin{align*}
{\bf D}^{\delta,\delta_1,n,{\cal S}}_{\bf k}
&:= {\bf D}^{\delta,n}_{\bf k} , \quad
 \forall {\bf k} \in 
Y_{\delta,\delta_1,n}({\cal S}) .
\end{align*}
As is proven in Appendixes \ref{asb} and \ref{asc},
the quantum variable-length source code
$({\bf E}^{\delta,\delta_1,n,{\cal S}} 
{\bf D}^{\delta,\delta_1,n,{\cal S}})$
on ${\cal H}^{\otimes n}$
satisfies
\begin{align}
&\epsilon_{n,p}({\bf E}^{\delta,\delta_1,n,{\cal S}},
{\bf D}^{\delta,\delta_1,n,{\cal S}})
\le 
1- 
\frac{C_{2,d}(n \delta_1)}{C_{1,d}(n \delta)}
\Bigl(
1- 
(n+d)^{4d }
\exp \left( -n 
C_{3,d} 
(\delta- \delta_1)^2
\right)
\Bigr)^{\frac{3}{2}} \Label{e12} \\
&\frac{-1}{n}\log
{\rm P}_{p^n}^{{\bf E}^{\delta,\delta_1,n,{\cal S}}}
\left\{ \frac{1}{n}
\left(\log | Y_{\delta,\delta_1,n,+}({\cal S})|
+ \log \dim {\cal H}_{{\bf k},n,\delta}
\right)
\ge R \right\} \nonumber \\
&\ge \frac{-5d}{n} \log (n+d) \nonumber \\
& \qquad +  
\min_{{\bf q}\in \real_+^{d,1}: H({\bf q}) \ge R - \frac{4d}{n}\log (n+d)
\exists \rho \in {\cal S} , \| {\bf q} - {\bf p}(\rho)
\| \le \delta_1}
\left( 
\min_{{\bf q}'\in \real_+^{d,1}: \| {\bf q} - {\bf q}' \|\le 2 \delta}
D( {\bf q}'\| {\bf p}) \right) 
\Label{e22},
\end{align}
for $ \forall p \in {\cal P}$.
Note that
$D({\rm p}(\rho) \|{\rm p}(\sigma))
= \min_{V:\hbox{\footnotesize unitary}}
D(\rho\|V^*\sigma V)$.
Letting $\delta:= n^{-1/4}$ and 
$\delta_1:= n^{-1/4} - n^{-1/3}$,
we can show that
the RHS of (\ref{e12}) tends to $0$, and that the RHS of (\ref{e22}) tends to the RHS of 
(\ref{op3}).

\section{Optimality of the exponent of the overflow probability}
\Label{s3} %:Proof of Theorem \ref{t1}
Next, we prove Theorem \ref{t1}.
When the support of any element $p$ of ${\cal P}$
consists of pure states, i.e. the pure states case,
we can prove Theorem \ref{t1} by using 
the monotonicity of quantum relative entropy
because the strong converse (\ref{D4-3}) holds
in quantum fixed-length pure state source coding,
as is explained in section \ref{s2.1}.
However, in the mixed states case,
we cannot use this strategy,
and we need the following lemma called 
the strong converse part of
quantum Stein's lemma in 
quantum hypothesis testing
proven by Ogawa and Nagaoka \cite{ON}
as an alternative.
Its another proof was given by Hayashi\cite{H2001}.
\begin{lem}\Label{L7}
Let $\rho$ and $\sigma$ be density operators on ${\cal H}$.
If any sequence of operators
$\vec{T}=\{ T_n \}$ on ${\cal H}^{\otimes n}$
satisfies 
that $0 \le T_n \le I$ and that
$
\liminf \Tr \rho^{\otimes n}T_n \,>0$,
then the inequality
\begin{align*}
\limsup\frac{-1}{n}\log \Tr \sigma^{\otimes n}T_n 
\le D( \rho\|\sigma)
\end{align*}
holds.
\end{lem}

Since the monotonicity
of quantum relative entropy corresponds
to the weak converse part of quantum Stein's lemma,
the former strategy can be regarded as
the combination of the strong converse part (\ref{D4-3})
of quantum fixed-length pure state source coding
and the weak converse part of
quantum Stein's lemma,
and the latter proof can be regarded as
 the combination of the weak converse part(\ref{D4})
of quantum fixed-length source coding
and the strong converse part of
quantum Stein's lemma.

First, for the reader's convenience,
we give the former proof which is 
simpler than the latter but is applied only in 
the pure states case.
After this proof, we give a more sound proof
which can be used in the general case.
Let $p$ and $q$ be an arbitrary elements of ${\cal P}$,
and $R$ be arbitrary real number such that
$R$ is less than the minimum admissible rate of $q$, i.e.,
$R \,< R_q$.
In particular, we assume that
the support of $q$ consists of pure states.
For a quantum variable-length source code
$\{({\bf E}^n,{\bf D}^n)\}$
for a family ${\cal P}$,
deforming the code $({\bf E}^n,{\bf D}^n)$,
we define the fixed-length code $(E^{R,n},D^{R,n})$
as follows.
When the data $\omega_n$ satisfies
\begin{align}
\log | \Omega_n | + \log \dim {\cal H}_{\omega_n}
\ge  n R, \Label{a1}
\end{align}
we send classical information
which indicates condition (\ref{a1}).
Otherwise, we send the data $\omega_n$ and 
the state $\frac{{\bf E}_{\omega_n}^n(\vec{\rho}_n)}
{\Tr {\bf E}_{\omega_n}^n(\vec{\rho}_n)}$.
In the decoding process,
if we receive the classical information
which indicates condition (\ref{a1}),
we regard a quantum state $\rho_R$
out of the original space ${\cal H}^{\otimes n}$
as the decoded state.
Note that $b(\rho_R,\rho)\le 1$ for any state 
$\rho \in {\cal S}({\cal H}^{\otimes n})$.
Otherwise,
we perform the operation ${\bf D}_{\omega_n}^n$
as the decoding process.
Since the maximum of this code is less than 
$nR$, we can regard it as a fixed-length code
whose length is $n R$.\par From the construction of the fixed-length code 
$(E^{R,n},D^{R,n})$,
we can easily check that
\begin{align*}
&\epsilon_{n,q}(E^{R,n},D^{R,n}) \\
\le &
{\rm P}_{q^n}^{{\bf E}^n}
\left\{ \frac{1}{n}
\left( \log | \Omega_n | + \log \dim {\cal H}_{\omega_n}
\right) \,< R \right\}
\epsilon_{n,q}
\left({\bf E}^n,{\bf D}^n\left |
 \frac{1}{n}
\left( \log | \Omega_n | + \log \dim {\cal H}_{\omega_n}
\right) \,< R 
\right.\right)\\
&+
{\rm P}_{q^n}^{{\bf E}^n}
\left\{ \frac{1}{n}
\left( \log | \Omega_n | + \log \dim {\cal H}_{\omega_n}
\right) \ge R \right\},
\end{align*}
where
$\epsilon_{n,q}
\left({\bf E}^n,{\bf D}^n\left |
\frac{1}{n}
\left( \log | \Omega_n | + \log \dim {\cal H}_{\omega_n}
\right) \,< R 
\right.\right)$
denotes the conditional average of the total error under the condition
$
\frac{1}{n}
\left(\log | \Omega_n | + \log \dim {\cal H}_{\omega_n}\right)
\,< R $.
Thus, we have the inequality
\begin{align}
& \epsilon_{n,q}(E^{R,n},D^{R,n})
- \epsilon_{n,q}({\bf E}^n,{\bf D}^n)\nonumber \\
\le &
{\rm P}_{q^n}^{{\bf E}^n}
\left\{ \frac{1}{n}
\left( \log | \Omega_n | + \log \dim {\cal H}_{\omega_n}
\right) \ge R \right\}
\left( 1- \epsilon_{n,q}
\left({\bf E}^n,{\bf D}^n \left |
\frac{1}{n}
\left( \log | \Omega_n | + \log \dim {\cal H}_{\omega_n}
\right) \ge R 
\right) \right.\right) \nonumber \\
 \le& {\rm P}_{q^n}^{{\bf E}^n}
\left\{ \frac{1}{n}
\left( \log | \Omega_n | + \log \dim {\cal H}_{\omega_n}
\right) \ge R \right\}
=
{\rm P}_{\overline{\rho}_q^{\otimes n}}^{{\bf E}^n}
\left\{ \frac{1}{n}
\left( \log | \Omega_n | + \log \dim {\cal H}_{\omega_n}
\right) \ge R \right\}. \Label{323-2}
\end{align}
Since the support of $q$ consists of pure states
and $H(\overline{\rho}_q)= R_q \,> R$,
we obtain the relation:
\begin{align*}
\epsilon_{n,q}(E^{R,n},D^{R,n})
\to 1 
\end{align*}
which is called the strong converse part of
the 
quantum fixed-length pure state source coding\cite{Winter}.
Since the universality guarantees the relation
\begin{align}
\epsilon_{n,q}({\bf E}^n,{\bf D}^n)
\to 0\Label{329},
\end{align}
we have
\begin{align}
{\rm P}_{n,q} := 
{\rm P}_{\overline{\rho}_q^{\otimes n}}^{{\bf E}^n}
\left\{ \frac{1}{n}
\left( \log | \Omega_n | + \log \dim {\cal H}_{\omega_n}
\right) \ge R \right\}
\to 1 \Label{323}.
\end{align}
Using the monotonicity
of quantum relative entropy,
we have
\begin{align*}
n D( \overline{\rho}_q \| \overline{\rho}_p)
=
D( \overline{\rho}_q ^{\otimes n}\| 
\overline{\rho}_p^{\otimes n})
\ge
{\rm P}_{n,q} 
\log \frac{{\rm P}_{n,q} }{{\rm P}_{n,p} }
+ \left( 1- {\rm P}_{n,q} \right)
\log \frac{1- {\rm P}_{n,q} }{1- {\rm P}_{n,p} },
\end{align*}
where
we define ${\rm P}_{n,p}$ similarly to (\ref{323}).
Since $- \left( 1- {\rm P}_{n,q} \right)
\log (1- {\rm P}_{n,p}) \ge 0$,
\begin{align*}
- \frac{\log {\rm P}_{n,p}}{n}
\le 
\frac{n D( \overline{\rho}_q \| \overline{\rho}_p)+
h({\rm P}_{n,q})}
{n {\rm P}_{n,q}} \to D( \overline{\rho}_q \| \overline{\rho}_p),
\end{align*}
where $h(x)$ is the binary entropy $
-x\log x - (1-x) \log (1-x)$.
Now, we obtain inequality
(\ref{op}) in the pure states case.

Next, we proceed the general case.
It follows from 
(\ref{D4}) and the inequality $R \,< R_q$ that
we have
\begin{align}
\liminf \epsilon_{n,q}(E^{R,n},D^{R,n})
\,> 0 \Label{329-1}.
\end{align} From (\ref{323-2}) and (\ref{329}),
the relation 
$
\liminf {\rm P}_{n,q} \,> 0$ holds.
There exists a POVM $M^n=\{M^n_{\omega_n}\}_{\omega_n}$
such that
\begin{align*}
\Tr \rho_n M^n_{\omega_n} =
\Tr {\bf E}^n_{\omega_n}(\rho_n), \quad
\forall \rho_n \in {\cal S}({\cal H}^{\otimes n}).
\end{align*}
Letting 
\begin{align*}
T_n:= \sum_{\omega_n: 
\frac{1}{n}
\left( \log | \Omega_n | + \log \dim {\cal H}_{\omega_n}
\right) \ge R }
M^n_{\omega_n},
\end{align*}
we have
${\rm P}_{n,q}= \Tr \overline{\rho}_q^{\otimes n}T_n$
and ${\rm P}_{n,p}= \Tr \overline{\rho}_p^{\otimes n}T_n$.
Thus, 
Lemma \ref{L7} guarantees that
\begin{align*}
\limsup - \frac{1}{n}\log {\rm P}_{n,p}
\le 
D( \overline{\rho}_q \| \overline{\rho}_p).
\end{align*}
Now, the proof is completed.
\section{Discussion}\Label{s6}
In our code,
the nonzero number $\delta$ is essential.
One may expect that
the quantum variable-length source code
$\{({\bf E}^{0,n}, {\bf D}^{0,n})\}$
is universal.
However, this code destroys
the input state by a quantum measurement
as follows.
\begin{lem}\Label{L1}
Assume that $d=2$ and
$\{| e_1 \rangle , | e_2 \rangle
\}$ is a CONS of $\complex^2$.
If the support of $p$ is pure states
$\{ | e_1 \rangle \langle e_1|,
| e_2 \rangle \langle e_2|\}$,
the average error 
$\epsilon_{n,p}({\bf E}^{0,n}, {\bf D}^{0,n})$ 
does not tends to $0$.
\end{lem}
As is understood from our proof of Theorem \ref{t2},
bound (\ref{op3}) cannot be achieved
unless $\delta$ tends to $0$.
It seems essential to
approximate the nonzero number $\delta \,> 0$
to $0$. 

If we discuss quantum universal coding under
 another error
$\epsilon_{n,p}'({\bf E}^n,{\bf D}^n)
$ instead of $\epsilon_{n,p}({\bf E}^n,{\bf D}^n)$
( c.f. (\ref{e90})):
\begin{align}
\epsilon_{n,p}'({\bf E}^n,{\bf D}^n)
& := 
\sum_{\vec{\rho}_n \in {\cal S}({\cal H})}p^n(\vec{\rho}_n)
\frac{1}{n}
\sum_{i=1}^n 
\sum_{\omega_n \in \Omega_n}
\Tr {\bf E}^n_{\omega_n}(\vec{\rho}_n)
b^2 \left(\rho_i, 
{\bf E}^n_{\omega_n}(\vec{\rho}_n)_i)\right) 
\Label{stst} \\
\vec{\rho}_n &= \rho_1 \otimes \cdots \otimes \rho_n 
\nonumber \\
{\bf E}^n_{\omega_n}(\vec{\rho}_n)_i &=
\Tr_{{\cal H}_1 \otimes \cdots \otimes {\cal H}_{i-1}
\otimes {\cal H}_{i+1}\otimes \cdots \otimes{\cal H}_{n}}
{\bf E}^n_{\omega_n}(\vec{\rho}_n),\nonumber
\end{align}
we can use several strategies for quantum universal coding.
For example, 
if we use $n \epsilon$ states only for the estimation
of $H(\overline{\rho}_p)$, 
we can reduce  
the error $\epsilon_{n,p}'$ to zero, asymptotically,
by use of Jozsa et. al. protocol \cite{JH}.
However, in this strategy,
we cannot reduce the error $\epsilon_{n,p}$
because the demolition of the first $n \epsilon$ states
is crucial for this criterion.

Next, we discuss
how rapidly the average error $\epsilon_{n,p}$ 
tends to $0$ in our code.
Assume that $d=2$ and
$\{| e_1 \rangle , | e_2 \rangle
\}$ is a CONS of $\complex^2$.
Unless $\delta_n \,> 0$ satisfies $|\delta _n| \,< 1$,
the coding length always
equals $2n$.
Then, we can assume that
$|\delta _n| \,< 1$.
\begin{lem}\Label{L2}
If the support of $p$ is pure states
$\{ | e_1 \rangle \langle e_1|,
| e_2 \rangle \langle e_2|\}$,
the relation 
\begin{align}
\lim \frac{-1}{n} \log 
\epsilon_{n,p}({\bf E}^{\delta_n,n},{\bf D}^{\delta_n,n})
= 0, \Label{13}
\end{align}
holds for any sequence $\{\delta_n\}$
satisfying $|\delta _n| \,< 1$.
\end{lem}
Therefore, it seems impossible
to construct a universal code
whose average error $\epsilon_{n,p}$ 
exponentially tends to $0$.

In general, even if 
$R_p= H(\overline{\rho}_p)$ for $\forall p
\in {\cal P}$,
the RHS of (\ref{op3}) does not necessarily
coincide with the RHS of (\ref{op}).
For example, when 
\begin{align*}
{\cal P}&= \{ p_t | t \in (0,1/2)\} , \quad
H(\overline{\rho}_{p_t})= R_{p_t}, \\
\overline{\rho}_{p_t}&=
\left( 
\begin{array}{cc}
t\cos^2 \theta(t) + (1-t)\sin^2 \theta(t) &
(1-2t) \cos \theta(t) \sin \theta(t) \\
(1-2t) \cos \theta(t) \sin \theta(t) &
(1-t)\cos^2 \theta(t) + t\sin^2 \theta(t) 
\end{array}
\right)
\end{align*}
and $\theta$ is continuous and one-to-one,
the both sides of (\ref{op9}) coincide with
$H(\overline{\rho}_{p_t})$
while the RHS of (\ref{op3}) is 
strictly smaller than the RHS of (\ref{op})
as follows.
For $t_1,t_0 \in (0,1/2)$, we can calculate as:
\begin{align*}
\inf_{t \in (0,1/2): H(\overline{\rho}_{p_t})
\ge h(t_1)}
\min_{V:unitary}
D(\overline{\rho}_{p_t}\| V
\overline{\rho}_{p_{t_0}}V^*) 
=& d(t_1, t_0) \\
\inf_{t \in (0,1/2): R_{p_t}
\,> h(t_1)}
D(\overline{\rho}_{p_t}\| 
\overline{\rho}_{p_{t_0}}) 
=&
\cos^2 (\theta(t)-\theta(t))
d(t_1, t_0) \\
&\quad + \sin^2 (\theta(t)-\theta(t))
d(t_1, 1-t_0)  .
\end{align*}
where
\begin{align*}
h(t) &:= - t\log t - (1-t)\log (1-t) \\
d(t,t')&:= t \log \frac{t}{t'} 
+ (1-t)\log \frac{1-t}{1-t'}.
\end{align*}
Thus, its difference 
equals $\sin^2 (\theta(t)-\theta(t))
\left( d(t_1, 1-t_0)  - d(t_1, t_0) \right)
\,> 0$.
This gap is closely related to the
ambiguity of the large deviation-type
bounds in quantum estimation \cite{H2002-2}.
It seems very hard to match the upper bound 
and the lower bound concerning the exponent of the
overflow probability in the general case.

\section{Conclusion}
We construct a quantum variable-length code
satisfying equation (\ref{op3}).
This is optimal 
in the sense of 
Theorem \ref{t1} when the family ${\cal P}$ 
consists 
of probabilities on ${\cal S}({\cal H})$
with no trivial redundancies.
However, in our code the average error does not 
exponentially vanish.
The construction of such a code seems to be 
difficult.

\section*{Acknowledgment}
The authors wish
to thank Professor H. Nagaoka and Dr. A. Winter 
for useful comments.
They also thank an anonymous referee for
useful comments.
\appendix
\section{Representation theoretical type method}
\Label{asa}
For our proof, we need the following two lemmas.
\begin{lem}
The relation
\begin{align}
%\left|\frac{1}{n}\log \dim {\cal V}_{{\bf n}} - H\left( \frac{{\bf n}}{n}
%\right)\right|
%& \le \frac{2d^2 +d}{2n}\log (n+d) +
%\frac{C}{n},\quad \forall {\bf n}.\Label{l10} \\
 \dim {\cal V}_{{\bf n}}
\le & C({\bf n})(n+d)^{2d}
\le (n+d)^{2d}
\exp \left( n H\left(\frac{\bf n}{n}\right)\right) \Label{a4} \\
\# \{ {\bf n} | {\bf n} \in Y_n\}
\le& (n+1)^d\Label{h20} \\
\dim {\cal U}_{\bf n} \le &(n+1)^d \Label{h21}
\end{align}
holds, where $C({\bf n})$ is defined as
\begin{align*}
C({\bf n}):= 
\frac{n!}{n_1! n_2! \ldots n_d !}. 
\end{align*}
\end{lem}
\begin{proof}
Inequality (\ref{h20})
is trivial. 
Using Young index ${\bf n}$,
the basis of ${\cal U}_{\bf n}$
is described by
$\{ e_{{\bf n}'} \}_{{\bf n}' \in Y^{\bf n}}$,
where $Y^{\bf n}$ is defined as
\begin{align*}
Y^{\bf n} 
:= \left\{
{\bf n}'= \{ n'_i\} \in \mathbb{Z}^d\left|
\begin{array}{l}
\sum_i n'_i=\sum_i n_i, \\
\sum_{i=1}^m n'_{s(i)} \le \sum_{i=1}^m
n_i, \\
\quad 1 \le \forall m \le d-1, \\
s \hbox{ is any permutation} 
\end{array}
\right.\right\}.
\end{align*}
Thus, we obtain (\ref{h21}).
Note that
the correspondence
${\bf n}'$ and $e_{{\bf n}'}$
depends on the choice of
Cartan subalgebra, i.e.
the choice of basis of ${\cal H}$.

According to Weyl \cite{Weyl}, Iwahori \cite{Iwa},
the following equation holds,
and it is evaluated as
\begin{align}
\dim {\cal V}_{{\bf n}}&=\frac{n !}{(n_1+d-1)! (n_2+d-2)! \ldots n_d!}
\prod_{j \,> i}(n_i-n_j-i+j)\nonumber \\
&\le \frac{n !}{n_1! n_2! \ldots n_d!}
\prod_{j \,> i}(n_i-n_j-i+j)
\le C({\bf n})(n+d)^{2d} \nonumber\\
&\le (n+d)^{2d}
\exp \left( n H\left(\frac{\bf n}{n}\right)\right). \nonumber 
\end{align}
\end{proof}
The following is essentially equivalent to
Keyl and Werner's result \cite{KW}.
For the reader's convenience, we give a simpler proof.
\begin{lem}\Label{lee}
Assume that
${\bf p}$ is the spectrum of $\rho$ such that
$p_1\ge p_2 \ge \ldots \ge p_d$.
The following relations
\begin{align}
\Tr P_{{\bf n}} \rho^{\otimes n}
&\le 
(n+d)^{3d }
\exp \left( -n 
D\left(\left. \frac{\bf n}{n} \right\|{\bf p}\right)\right)
\Label{e31} \\
\sum_{
\frac{\bf n}{n} \notin {\cal R}}
\Tr P_{{\bf n}} \rho^{\otimes n}
&\le
(n+d)^{4d }
\exp \left( -n 
\min_{q \notin {\cal R}}
D( {\bf q} \|{\bf p})
\right)
\Label{e32}
\end{align}
hold for a subset ${\cal R}$ of $\real_+^{d,1}$.
\end{lem}
\begin{proof}
Let ${\cal U}_{\bf n}'$ be an irreducible
representation of $SU(d)$ in ${\cal H}^{\otimes n}$,
which is equivalent to ${\cal U}_{\bf n}$.
We denote its projection by $P_{\bf n}'$.
Now, we choose the basis 
$\{ e_{{\bf n}'}\}_{{\bf n}' \in Y^{\bf n}}$
of ${\cal U}_{\bf n}'$
depending the basis $\{e_i\}$ of ${\cal H}$.
The base $e_{{\bf n}'}$ is the eigenvector 
of $\rho^{\otimes n}$ with the
eigenvalue $\prod_{i=1}^d a_i^{n_i'}$.
Since ${\bf n}'$ is majorized by ${\bf n}$,
we can calculate the operator norm by
\begin{align}
\left \|
P_{\bf n}' \rho^{\otimes n}
P_{\bf n}' 
\right\| = \prod_{i=1}^d a_i^{n_i},\Label{a5}
\end{align}
where $\| X \| := \sup_{x \in {\cal H}}\| X x \|$.
from (\ref{a4}), (\ref{h21}) and (\ref{a5}), the relations
\begin{align*}
\Tr P_{{\bf n}} \rho^{\otimes n}
&=
\dim {\cal V}_{\bf n} \times 
\Tr
P_{\bf n}' \rho^{\otimes n}
\le  
(n+d)^{3d} C({\bf n}) \prod_{i=1}^d a_i^{n_i}\\
&= (n+d)^{3d} {\rm Mul}( {\bf a}, {\bf n} )  
\end{align*}
hold, where 
we denote the multinomial distribution of ${\bf a}$ by
${\rm Mul}( {\bf a}, \bullet )$.
Thus, we obtain (\ref{e31}).
Inequality (\ref{h20}) guarantees 
\begin{align*}
\sum_{
\frac{\bf n}{n} \notin {\cal R}}
\Tr P_{{\bf n}} \rho^{\otimes n}
\le
(n+d)^{4d}
\exp \left( -n 
\inf_{{\bf q} \notin {\cal R}}
D( {\bf q} \|{\bf p})
\right).
\end{align*}
\end{proof}

\section{Proof of (\ref{e2}) and (\ref{e22})}\Label{asb}
First, we prove inequality (\ref{e2}).
For a sufficiently large integer $n$,
the relations 
\begin{align*}
| Y_{\delta,n}|
\le \# \{ {\bf k} \in \integer^d
| k_i \ge 0 \}
\le (n+1)^d
\end{align*}
hold.
Since 
$\dim {\cal U}_{\bf n}
\le (n+d)^d$,
for any ${\bf k} \in Y_{\delta,n}$, we have
\begin{align*}
\log | Y_{\delta,n} | + \log \dim 
{\cal H}_{\bf k}^{\delta, n}
&\le
d \log (n+1) + 
\max_{{\bf n}\in Y_{\delta,n}\cap U_{{\bf k},n \delta}}
\log \dim {\cal U}_{\bf n} + \log \dim {\cal V}_{\bf n}  \\
&\le
4d \log (n+d)
+ \max_{{\bf n}\in Y_{\delta,n}\cap U_{{\bf k},n \delta}}
n H \left( \frac{\bf n}{n} \right).
\end{align*} From (\ref{e32}),
we have
\begin{align*}
\Tr M_{\bf k}^{\delta ,n} \overline{\rho}_p^{\otimes n}
&\le
\frac{|Y_{\delta,n}|}{C_{1,d}(n \delta)}
(n+d)^{3d}
\max_{{\bf n}' \in Y_n \cap U_{{\bf k},n\delta}}
\exp \left(
- n D \left(\left. \frac{{\bf n}'}{n} \right\|
{\bf p}\right)
\right) \\
&\le
(n+d)^{4d}
\max_{{\bf n}' \in Y_n \cap U_{{\bf k},n\delta}}
\exp \left(
- n D \left(\left. \frac{{\bf n}'}{n} \right\|
{\bf p}\right)
\right) \\
 &\le
(n+d)^{4d}
\max_{{\bf q} \in U_{\frac{\bf k}{n},\delta} \cap \real_+^
{d,1}}
\exp \left(
- n D \left(\left. {\bf q} \right\|
{\bf p}\right)
\right)  .
\end{align*}
Thus,
\begin{align*}
%&\frac{-1}{n}\log 
&{\rm P}_{\overline{\rho}_{p}^{\otimes n}}^{{\bf E}^n}
\left\{ \frac{1}{n}
\left(\log | Y_{\delta,n} | 
+ \log \dim {\cal H}_{\bf k}^{\delta, n}
\right)
\ge R \right\} \\
&\le
\sum_{
{\bf k}\in Y_{\delta,n}: {\displaystyle\max_{{\bf n}\in Y_n
\cap U_{{\bf k},n \delta}}}
H(\frac{\bf n}{n}) \ge R - \frac{4d}{n}\log (n+d)}
\Tr M_{\bf k}^{\delta, n} \overline{\rho}_p^{\otimes n}\\
& \le
|Y_{\delta,n}|(n+d)^{4d}
\max_{{\bf n}\in Y_n: H(\frac{\bf n}{n}) \ge R - \frac{4d}{n}\log (n+d)}
\max_{{\bf n}'\in Y_n: \| {\bf n} - {\bf n}' \|\le 2 \delta n}
\exp \left(
- n D \left(\left. \frac{{\bf n}'}{n} \right\|
{\bf p}\right)
\right) \\
& \le
(n+d)^{5d}
\max_{{\bf q}\in \real_+^{d,1}: H({\bf q}) \ge R - \frac{4d}{n}\log (n+d)}
\max_{{\bf q}'\in \real_+^{d,1}: \| {\bf q} - {\bf q}' \|\le 2 \delta}
\exp \left(
- n D \left(\left. {\bf q}' \right\|
{\bf p}\right)
\right) .
\end{align*}
Then, we obtain (\ref{e2}).

Next, we proceed to (\ref{e22}).
Since 
$|Y_{\delta,\delta_1,n,+}({\cal S})|
\le | Y_{\delta,n}|$,
we have
\begin{align*}
%&\frac{-1}{n}\log 
&{\rm P}_{\overline{\rho}_{p}^{\otimes n}}^{{\bf E}^n}
\left\{ \frac{1}{n}
\left(\log | Y_{\delta,\delta_1,n,+}({\cal S})|
+ \log \dim {\cal H}_{\bf k}^{\delta, n}
\right)
\ge R \right\} \\
&\le
\sum_{
{\bf k}\in Y_{\delta,\delta_1,n,+}({\cal S})
: {\displaystyle
\max_{{\bf n}\in Y_n \cap U_{{\bf k},n \delta}}}
H(\frac{\bf n}{n}) \ge R - \frac{4d}{n}\log (n+d)}
\Tr  M_{\bf k}^{\delta, n} \overline{\rho}_p^{\otimes n}\\
& \le
|Y_{\delta,n}|(n+d)^{4d}
\max_{{\bf n}\in Y_n: H(\frac{\bf n}{n}) \ge R - \frac{4d}{n}\log (n+d),
\exists \rho \in {\cal S} , \| \frac{\bf n}{n} - {\bf p}(\rho)
\| \le \delta_1 }
\max_{{\bf n}'\in Y_n: \| {\bf n} - {\bf n}' \|\le 2 \delta n}
\exp \left(
- n D \left(\left. \frac{{\bf n}'}{n} \right\|
{\bf p}\right)
\right) \\
& \le
(n+d)^{5d}
\max_{{\bf q}\in \real_+^{d,1}: H({\bf q}) \ge R - \frac{4d}{n}\log (n+d)
\exists \rho \in {\cal S} , \| {\bf q} - {\bf p}(\rho)
\| \le \delta_1}
\max_{{\bf q}'\in \real_+^{d,1}: \| {\bf q} - {\bf q}' \|\le 2 \delta}
\exp \left(
- n D \left(\left. {\bf q}' \right\|
{\bf p}\right)
\right) .
\end{align*}
Then, we obtain (\ref{e22}).

\section{Proof of (\ref{e1}), (\ref{e1-3}) and (\ref{e12})}\Label{asc}
We can evaluate the average error as
\begin{align}
&\epsilon_{n,p}({\bf E}^{\delta,n},{\bf D}^{\delta,n}) 
\nonumber \\
& =\sum_{\vec{\rho}_n \in {\cal S}({\cal H}^{\otimes n})}
p^n(\vec{\rho}_n)
\sum_{{\bf k} \in Y_{\delta,n}}
\Tr M_{\bf k}^{\delta, n} \vec{\rho}_n
\left( 1- \Tr \left|\sqrt{\vec{\rho}_n}
\sqrt{\frac{\sqrt{M_{\bf k}^{\delta, n}}
\vec{\rho}_n
\sqrt{M_{\bf k}^{\delta, n}}}
{\Tr  M_{\bf k}^{\delta, n}\vec{\rho}_n}
}\right|\right) \nonumber \\
& =
1- \sum_{\vec{\rho}_n \in {\cal S}({\cal H}^{\otimes n})}
p^n(\vec{\rho}_n)
\sum_{{\bf k} \in Y_{\delta,n}}
\sqrt{\Tr M_{\bf k}^{\delta, n} \vec{\rho}_n}
\Tr \sqrt{\sqrt{\vec{\rho}_n}
\sqrt{M_{\bf k}^{\delta, n}}
\vec{\rho}_n
\sqrt{M_{\bf k}^{\delta, n}}
\sqrt{\vec{\rho}_n}} \nonumber\\
& =
1- \sum_{\vec{\rho}_n \in {\cal S}({\cal H}^{\otimes n})}
p^n(\vec{\rho}_n)
\sum_{{\bf k} \in Y_{\delta,n}}
\sqrt{\Tr M_{\bf k}^{\delta, n} \vec{\rho}_n}
\Tr \sqrt{\vec{\rho}_n}
\sqrt{M_{\bf k}^{\delta, n}}
\sqrt{\vec{\rho}_n} \nonumber\\
& =
1- 
\sum_{{\bf k} \in Y_{\delta,n}}
\frac{1}{C_{1,d}(n \delta)}
\sum_{\vec{\rho}_n \in {\cal S}({\cal H}^{\otimes n})}
p^n(\vec{\rho}_n)
\left(\Tr P_{\bf k}^{\delta, n} \vec{\rho}_n
\right)^{\frac{3}{2}}
\nonumber\\
& \le
1- \sum_{{\bf k} \in Y_{\delta,n}}
\frac{1}{C_{1,d}(n \delta)}
\left(
\sum_{\vec{\rho}_n \in {\cal S}({\cal H}^{\otimes n})}
p^n(\vec{\rho}_n)\Tr P_{\bf k}^{\delta, n} \vec{\rho}_n
\right)^{\frac{3}{2}} \Label{jensen} \\
& =
1- \sum_{{\bf k} \in Y_{\delta,n}}
\frac{1}{C_{1,d}(n \delta)}
\left( \Tr \overline{\rho}_p^{\otimes n}
P_{\bf k}^{\delta, n}
\right)^{\frac{3}{2}} ,\Label{e51}
\end{align}
where inequality (\ref{jensen}) follows from 
Jensen's inequality concerning 
the convex function $x \mapsto x^{3/2}$.

The relations 
\begin{align}
C_{2,d}(n \delta_1)
&\le \# \left(Y_{\delta,n} \cap U_{n {\bf p},n \delta_1}\right),
\quad 0 \,< \delta_1 \,< \delta
\Label{e54} \\
P_{\bf k}^{\delta, n}
& \ge
\sum_{{\bf n} \in Y_n \cap U_{n{\bf p},n (\delta - \delta_1)}}
P_{\bf n} , \quad
\forall {\bf k} \in 
Y_{\delta,n} \cap U_{n {\bf p},n \delta_1}\Label{e52}
\end{align}
hold.
Using Lemma \ref{lee}, and equations (\ref{e52}) and (\ref{a3}), we have
\begin{align}
\Tr 
P_{\bf k}^{\delta, n}
\overline{\rho}_p^{\otimes n}
&\ge
1- 
(n+d)^{4d }
\exp \left( -n 
\min_{q \notin U_{{\bf p},\delta- \delta_1 }}
D( {\bf q} \|{\bf p})
\right) \nonumber \\
&\ge
1- 
(n+d)^{4d }
\exp \left( -n C_{3,d} (\delta- \delta_1 )^2\right).
\Label{e53}
\end{align}
It follows from (\ref{e54}) and (\ref{e53}) that
\begin{align}
\sum_{{\bf k} \in Y_{\delta,n}}
\frac{1}{C_{1,d}(n \delta)}
\left( \Tr \overline{\rho}_p^{\otimes n}
P_{\bf k}^{\delta, n}
\right)^{\frac{3}{2}} 
&\ge
\frac{1}{C_{1,d}(n \delta)}
\sum_{{\bf k} \in 
 Y_{\delta,n} \cap U_{n{\bf p},n \delta_1}}
\left( \Tr \overline{\rho}_p^{\otimes n}
P_{\bf k}^{\delta, n}
\right)^{\frac{3}{2}} \nonumber \\
& \ge
\frac{C_{2,d}(n \delta_1)}{C_{1,d}(n \delta)}
\Bigl(
1- 
(n+d)^{4d }
\exp \left( -n 
C_{3,d} 
(\delta- \delta_1)^2
\right)
\Bigr)^{\frac{3}{2}}. \Label{e55}
\end{align}
Inequality (\ref{e1}) follows from (\ref{e51})
and (\ref{e55}).

Similarly to (\ref{e51}), we can prove 
\begin{align*}
&\epsilon_{n,p}''({\bf E}^{\delta,n},{\bf D}^{\delta,n}) 
\le 1- \sum_{{\bf k} \in Y_{\delta,n}}
\frac{1}{C_{1,d}(n \delta)}
\left( \Tr \overline{\rho}_p^{\otimes n}
P_{\bf k}^{\delta, n}
\right)^{2} ,
\end{align*}
which implies (\ref{e1-3}).

In the general case, similarly to (\ref{e51}),
we can prove that
\begin{align}
\epsilon_{n,p}({\bf E}^{\delta,n},{\bf D}^{\delta,n}) 
 \le
1- \sum_{{\bf k} \in Y_{\delta,\delta_1,n}({\cal S})}
\frac{1}{C_{1,d}(n \delta)}
\left( \Tr \overline{\rho}_p^{\otimes n}
P_{\bf k}^{\delta, n}
\right)^{\frac{3}{2}} . \Label{e56}
\end{align}
Since $
Y_{\delta,\delta_1,n,+}({\cal S})
\cap U_{n{\bf p},n \delta_1}=
 Y_{\delta,n} \cap U_{n{\bf p},n \delta_1}$,
we can prove that
\begin{align}
& \sum_{{\bf k} \in Y_{\delta,\delta_1,n,+}({\cal S})}
\frac{1}{C_{1,d}(n \delta)}
\left( \Tr \overline{\rho}_p^{\otimes n}
P_{\bf k}^{\delta, n}
\right)^{\frac{3}{2}} \nonumber \\
&\ge
\frac{1}{C_{1,d}(n \delta)}
\sum_{{\bf k} \in 
 Y_{\delta,n} \cap U_{n{\bf p},n \delta_1}}
\left( \Tr \overline{\rho}_p^{\otimes n}
P_{\bf k}^{\delta, n}
\right)^{\frac{3}{2}} \nonumber \\
& \ge
\frac{C_{2,d}(n \delta_1)}{C_{1,d}(n \delta)}
\Bigl(
1- 
(n+d)^{4d }
\exp \left( -n 
C_{3,d} 
(\delta- \delta_1)^2
\right)
\Bigr)^{\frac{3}{2}}. \Label{e58}
\end{align}
Inequality (\ref{e12}) follows from (\ref{e56})
and (\ref{e58}).

\section{Proof of Lemma \ref{L1}}
In this case,
the average error is 
calculated as
\begin{align*}
\epsilon_{n,p}({\bf E}^{0,n},{\bf D}^{0,n})
&= 1- \sum_{{\bf n} \in Y_n}
\sum_{\vec{e}_n} p(\vec{e}_n)
\left(
\langle \vec{e}_n | P_{\bf n} | \vec{e}_n \rangle
\right)^{\frac{3}{2}} \\
&= 1- \sum_{\vec{e}_n} p(\vec{e}_n)
\sum_{{\bf n} \in Y_n}
\left(
\langle \vec{e}_n | P_{\bf n} | \vec{e}_n \rangle
\right)^{\frac{3}{2}},
\end{align*}
where $\vec{e}_n:= e_{i_1} \otimes e_{i_2}
\otimes \cdots \otimes e_{i_n} \in
{\cal H}^{\otimes n}$.
We define 
${\bf n}(\vec{e}_n):= ( n_1(\vec{e}_n), n_2(\vec{e}_n))$ 
by
\begin{align}
n_i(\vec{e}_n) := \# \{ j \in [ 1, n] | e_{i_j}= e_i\}.
\Label{e61}
\end{align}
Now, we focus a typical element $\vec{e}_n$,
i.e. $\frac{n_i(\vec{e}_n)}{n} \cong p_i$.
The number satisfying (\ref{e61})
is ${n(\vec{e}_n) \choose n_2(\vec{e}_n) }$, and 
$\dim {\cal V}_{{\bf n}'}= 
\left({n \choose n_2(\vec{e}_n) }- {n \choose n_2(\vec{e}_n) -1 }\right)$,
where ${\bf n}(\vec{e}_n)=(n_1(\vec{e}_n), n_2(\vec{e}_n)) \in Y_n$.
Then, 
\begin{align*}
\langle \vec{e}_n | P_{{\bf n}(\vec{e}_n)} | \vec{e}_n \rangle
& = {n \choose n_2(\vec{e}_n) }^{-1}
\left({n \choose n_2(\vec{e}_n) }- {n \choose n_2(\vec{e}_n) -1 }\right) \\
%= \frac{n_1 ! n_2 !}{n !}
%\left( 
%\frac{n !}{n_1 ! n_2 !}- \frac{n !}{(n_1+1) ! (n_2-1) !}
%\right)\\
& = 1- \frac{n_2(\vec{e}_n)}{n_1(\vec{e}_n)+1} =
%\frac{n_1 + 1 - n_2}{n_1+1}=
\frac{\frac{n_1(\vec{e}_n)}{n} + \frac{1}{n} - \frac{n_2(\vec{e}_n)}{n}}
{\frac{n_1(\vec{e}_n)}{n} + \frac{1}{n}}
\cong \frac{p_1-p_2}{p_1}.
\end{align*}
Since $x^{\frac{3}{2}}+ y^{\frac{3}{2}}
\le (x+y)^{\frac{3}{2}}$ for $0\,< x,y \,<1$,
we can evaluate
\begin{align*}
\sum_{{\bf n} \in Y_n}
\left(
\langle \vec{e}_n | P_{\bf n} | \vec{e}_n \rangle
\right)^{\frac{3}{2}}
&\le
\left(
\sum_{{\bf n} \in Y_n \setminus \{{\bf n}'\} }
\langle \vec{e}_n | P_{\bf n} | \vec{e}_n \rangle
\right)^{\frac{3}{2}}
+
\left(\langle \vec{e}_n | P_{\bf n}' | \vec{e}_n \rangle
\right)^{\frac{3}{2}} \\
&\cong 
\left(
1- \frac{p_1-p_2}{p_1}
\right)^{\frac{3}{2}}
+
\left(\frac{p_1-p_2}{p_1}
\right)^{\frac{3}{2}} \\
& = 
\left(
\frac{p_2}{p_1}
\right)^{\frac{3}{2}}
+
\left(\frac{p_1-p_2}{p_1}
\right)^{\frac{3}{2}} \,< 1.
\end{align*}
Therefore,
\begin{align*}
\lim \epsilon_{n,p}({\bf E}^{0,n},{\bf D}^{0,n})
&\ge
1- \left( \left(
\frac{p_2}{p_1}
\right)^{\frac{3}{2}}
+
\left(\frac{p_1-p_2}{p_1}
\right)^{\frac{3}{2}}\right)
\,> 0 .
\end{align*}
\section{Proof of Lemma \ref{L2}}
For any ${\bf n}\in Y_n, \delta_n \,> 0$,
we denote 
$([n_1- \frac{1}{\sqrt{2}} \delta_n] , 
n- [n_1 -\frac{1}{\sqrt{2}} \delta_n] )
\in Y_{\delta,n}$
by ${\bf k}({\bf n},\delta_n)$,
where
$[x]$ is defined as the maximum integer
$n$ satisfying $n \le x$.
The element ${\bf k}({\bf n},\delta_n)$
satisfies
\begin{align*}
{\bf n}=(n_1,n_2) \in U_{{\bf k}({\bf n},\delta_n),\delta_n} \\
(n_1+1,n_2-1) 
\notin U_{{\bf k}({\bf n},\delta_n),\delta_n} .
\end{align*}
For any $\delta\,> 0$, we have
\begin{align}
&\epsilon_{n,p}({\bf E}^{\delta,n},{\bf D}^{\delta,n}) 
\nonumber \\
& =
\sum_{\vec{e}_n}p^n(\vec{e}_n)
\left( 1- 
\sum_{{\bf k} \in Y_{\delta,n}}
\frac{1}{C_{1,d}(n \delta)}
(\Tr P_{\bf k}^{\delta, n} \vec{\rho}_n)^{\frac{3}{2}}
\right)
\nonumber\\
& \ge
\sum_{\vec{e}_n}p^n(\vec{e}_n)
\left( 1- 
\sum_{{\bf k} \neq {\bf k}({\bf n},\delta_n)
\in Y_{\delta,n}}
\frac{1}{C_{1,d}(n \delta)}
(\Tr P_{{\bf k}}^{\delta, n} \vec{\rho}_n)
-
\frac{1}{C_{1,d}(n \delta)}
(\Tr P_{{\bf k}({\bf n},\delta_n)}^{\delta, n} \vec{\rho}_n)^{\frac{3}{2}}
\right)
\nonumber\\
& =
\sum_{\vec{e}_n}p^n(\vec{e}_n)
\left( 
\frac{1}{C_{1,d}(n \delta)}
(\Tr P_{{\bf k}({\bf n},\delta_n)}^{\delta, n} \vec{\rho}_n)
-
\frac{1}{C_{1,d}(n \delta)}
(\Tr P_{{\bf k}({\bf n},\delta_n)}^{\delta, n} \vec{\rho}_n)^{\frac{3}{2}}
\right)
\nonumber\\
& =
\sum_{\vec{e}_n}p^n(\vec{e}_n)
\frac{1}{C_{1,d}(n \delta)}
\left (
\Tr P_{{\bf k}({\bf n},\delta_n)}^{\delta, n} \vec{\rho}_n
-
(\Tr P_{{\bf k}({\bf n},\delta_n)}^{\delta, n} \vec{\rho}_n)^{\frac{3}{2}}
\right)
\nonumber\\
& \ge
\sum_{\vec{e}_n : 
\frac{n_i(\vec{e}_n)}{n} \cong p_i}
p^n(\vec{e}_n)
\frac{1}{C_{1,d}(n \delta)}
\left (
\Tr P_{{\bf k}({\bf n},\delta_n)}^{\delta, n} \vec{\rho}_n
-
(\Tr P_{{\bf k}({\bf n},\delta_n)}^{\delta, n} \vec{\rho}_n)^{\frac{3}{2}}
\right)
\nonumber\\
& \cong
\sum_{\vec{e}_n : 
\frac{n_i(\vec{e}_n)}{n} \cong p_i}
p^n(\vec{e}_n)
\frac{1}{C_{1,d}(n \delta)}
\left (
\frac{p_1-p_2}{p_1}
-
\left(\frac{p_1-p_2}{p_1}\right)^{\frac{3}{2}}
\right)
\nonumber\\
& \cong
\frac{1}{C_{1,d}(n \delta)}
\left (
\frac{p_1-p_2}{p_1}
-
\left(\frac{p_1-p_2}{p_1}\right)^{\frac{3}{2}}
\right)
\ge
\frac{1}{2|Y_n|}
\left (
\frac{p_1-p_2}{p_1}
-
\left(\frac{p_1-p_2}{p_1}\right)^{\frac{3}{2}}
\right)
.\nonumber
\end{align}
Note that the RHS is independent of $\delta\,>0$.
Thus,
\begin{align*}
\frac{-1}{n}
\log \epsilon_{n,p}({\bf E}^{\delta,n},{\bf D}^{\delta,n}) 
&\le
\frac{-1}{n}
\left(
\log \frac{1}{2 |Y_n|}+
\log \left (
\frac{p_1-p_2}{p_1}
-
\left(\frac{p_1-p_2}{p_1}\right)^{\frac{3}{2}}
\right)
\right) \\
&\le
\frac{1}{n}
\left( \log 2 (n+1)^2
- \log \left (
\frac{p_1-p_2}{p_1}
-
\left(\frac{p_1-p_2}{p_1}\right)^{\frac{3}{2}}
\right)
\right)
\to 0.
\end{align*}
Therefore, 
we obtain (\ref{13}).

\end{document}